\documentclass[10pt, conference]{IEEEtran}
\ifCLASSINFOpdf
  \usepackage[pdftex]{graphicx}
  \graphicspath{{images/}}
  \DeclareGraphicsExtensions{.pdf,.jpeg,.png}
\else
  \usepackage[dvips]{graphicx}
  \graphicspath{{../eps/}}
  \DeclareGraphicsExtensions{.eps}
\fi
\usepackage{url}

\usepackage{caption}
\usepackage{hyperref}
\hypersetup{
   pdfborder={0 0 0},
   pdftitle={Built to Last or Built Too Fast? Evaluating Prediction Models for Build Times},
   pdfauthor={Ekaba, Eric, Olga},
   pdfsubject={Prepared for MSR Mining Challenge 2017},
   pdfkeywords={},
   pdfproducer={},
   pdfcreator={},
   colorlinks=true,
   citecolor=black,
   filecolor=black,
   linkcolor=black,
   urlcolor=black
}

\begin{document}
%
\title{Built to Last or Built Too Fast? Evaluating Prediction Models for Build Times}


\author{\IEEEauthorblockN{Ekaba Bisong, Eric Tran, Olga Baysal}
\IEEEauthorblockA{Department of Computer Science\\
Carleton University, Ottawa, Canada\\
ekaba.bisong@carleton.ca,\\
eric.tran@carleton.ca,\\
olga.baysal@carleton.ca}
}


%


\maketitle

\begin{abstract}
Automated builds are integral to the Continuous
Integration (CI) software development practice. In CI, developers
are encouraged to integrate early and often. However, long build
times can be an issue when integrations are frequent. This
research focuses on finding a balance between integrating often
and keeping developers productive. We propose and analyze
models that can predict the build time of a job. Such models
can help developers to better manage their time and tasks. Also,
project managers can explore different factors to determine the
best setup for a build job that will keep the build wait time
to an acceptable level. Software organizations transitioning to
CI practices can use the predictive models to anticipate build
times before CI is implemented. The research community can
modify our predictive models to further understand the factors
and relationships affecting build times.

\end{abstract}


%
\IEEEpeerreviewmaketitle

\section{Introduction}

Automated builds are integral to the Continuous Integration (CI) software development practice. As developers check in code to the shared repository, an automated system picks up the changes and triggers a build. The automated build will compile the code, and ideally, run a test suite. Build results notify developers about integration problems like compilation errors or missing dependencies. When combined with unit tests, build results can reveal broken or changed functionality in a software project.

In CI, developers are encouraged to commit their changes early and often. Changes that are smaller and more regularly integrated are easier to debug when something breaks \cite{meyer:cianditstools}. Thus, build times are very important when integrations are frequent. Long build times can become a bottleneck to the CI process.

Long build times are problematic for developers. Developers can lose focus and productivity while waiting for a build to finish. For example, developers may work on specific tasks in separate branches using a version control system. Should a build fail, it would be cumbersome to switch back to the original branch due to factors such as caching, configuration files, and so on. Additionally, developers may experience context switching when changing between different tasks. The cost of context switching can be low when the complexity of tasks is low. Conversely, when the complexity of tasks is high, context switching can be a costly expenditure of mental energy on the part of the developer \cite{kai:context}. Therefore, it may be easier to stay in the current branch and wait for the build results to finish before moving on to a future task or continuing with the current job.

We are interested in finding a balance between integrating often and keeping developers productive. Our research goal is simple: to build a predictive model that can predict the estimated build time of a job. Our project takes advantage of TravisTorrent, a freely available dataset combining features from GitHub and Travis CI for builds of more than 1,000 projects. 

\section{Related Work}

Mokhov et al. \cite{mokhov:make} noted that build systems face issues when projects have a large number of components, multiple languages, and complex interdependencies. For example, build systems may have executables that depend on libraries or running tools that are generated by the same build system. As well, build systems can be hard to maintain when build environments are not managed. Employing tools such as the Nix package manager can help describe package build actions and their dependencies, allowing the build environments to be produced automatically \cite{dolstra:nix}.

The effects of long build times can be negative. Brooks \cite{brooks:teampace} noted that long build times can affect the following variables: commit size, commit frequency, build down time, development flow, and developer satisfaction. Negative perceptions about waiting times can be lowered by providing feedback, controlling perceived waiting time, and having different waiting times for different tasks \cite{laukkanen:buildwait}.

As a solution to slow builds, Ammons \cite{ammons:scriptedbuilds} suggested breaking large, all-or-nothing builds into many  smaller builds. To demonstrate the technique, Ammons implemented a tool suite called Grexmk. Grexmk contains tools for dividing large builds into mini-builds and tools for executing mini-builds in parallel and incrementally. Moreover, a mini-build is an all-or-nothing build that explicitly lists its: output files, source files, dependences on other mini-builds, and build script. Overall, incremental builds were sped up by a factor of 1.2 when using Grexmk.

Brooks \cite{brooks:teampace} suggested that a build time of 2 minutes was optimal; build times under 10 minutes were considered acceptable. However, the suggested build times were based on experience reports from the same company. In summary, there is a lack of empirical quantitative research to address optimal build times in a CI environment \cite{laukkanen:buildwait}.

\section{Methodology}
\subsection{Dataset}
We selected TravisTorrent as the dataset for our prediction task. TravisTorrent is a synthesis of pull-request commits from software projects hosted on GitHub with TravisCI integrated as a mechanism for continuous integration \cite{msr17challenge}. The dataset draws features from GitHub for a particular build job in a project as well as corresponding build features from TravisCI. The data contains over 2 million records spanning 1,000 projects.

Our paper uses this information to investigate the factors that affect the build time of a build job in a pull-request CI development ecosystem. As well, we will build a predictive model to estimate the build time of a particular build job given that a specified set of build job features is known.

\subsection{Response Feature}
TravisTorrent has a feature called \textit{tr\_duration} which is a vector containing the overall duration of the build in seconds. This includes both the time taken for the Travis build to start, the time taken to run tests, and finally the time it takes to run the build. We selected \textit{tr\_duration} as the response variable for prediction because this is the estimated total time the developer is inert as he waits for the build to complete.

\subsection{Initial Data Preparation}
TravisTorrent contains 2,640,825 build records with 56 features. We first carry-out some rudimentary data preparation operations to get our data ready for analysis. The data is randomized to eliminate ordering in the data. Next we removed the records that contained no values (i.e. NA's) for the overall build duration (\textit{tr\_duration}) feature. We then create a 70\% to 30\% split of our data into a training/cv set and test set (this would later be used to evaluate the model to ascertain generalizability to unseen sample points). The training/cv set contains 1,846,396 records while the test set contains 791,310 records.

\subsection{Initial Feature Selection}
The dataset contains 56 feature vectors, of which 34 are integers/numeric, 16 are strings, 4 are booleans, and 2 are in the ISO date format. To begin building our predictive model, we selected all but 3 of the integer/numeric features. The features excluded were: unique records which indicated the unique identifier for each build job, the pull request number on GitHub, and the build number on Travis. We did not deem any of the string and date variables to be useful enough to be included as features in our predictive model. Finally, we considered all the boolean variables (which will be coded as factors) in this initial feature selection phase. This totaled 35 features selected for use in building the predictive model from the original 56 features. Table \ref{selected-features} lists the features considered for the prediction task.

\begin{table*}[t]
	\centering
	\caption{Selected Features for Prediction Model}
	\label{selected-features}
	\begin{tabular}{|l|p{13cm}|}
		\hline
		\bfseries{Feature} & \bfseries{Description} \\ 
		\hline
		gh\_team\_size&Number of developers that committed directly or merged pull requests from the moment the build was triggered and 3 months back  \\ 
		\hline 
		gh\_num\_issue\_comments&If git\_commit is linked to a PR on GitHub, the number of discussion comments on that PR  \\ 
		\hline 
		gh\_num\_commit\_comments&The number of comments on git\_commits on GitHub  \\ 
		\hline 
		gh\_num\_pr\_comments&The number of comments (code review) on this pull request on GitHub  \\ 
		\hline 
		gh\_src\_churn&How much (lines) production code changed in the commits built by this build  \\ 
		\hline 
		gh\_test\_churn&How much (lines) test code changed in the commits built by this build  \\ 
		\hline 
		gh\_files\_added&Number of files added by the commits built by this build  \\ 
		\hline 
		gh\_files\_deleted&Number of files deleted by the commits built by this build  \\ 
		\hline 
		gh\_files\_modified&Number of files modified by the commits built by this build  \\ 
		\hline 
		gh\_tests\_added& Lines of testing code added by the commits built by this build \\ 
		\hline 
		gh\_tests\_deleted&Lines of testing code deleted by the commits built by this build  \\ 
		\hline 
		gh\_src\_files&Number of src files changed by the commits that where built  \\ 
		\hline 
		gh\_doc\_files&Number of documentation files changed by the commits that where built  \\ 
		\hline 
		gh\_other\_files&Number of files which are neither production code nor documentation that changed by the commits that where built  \\ 
		\hline 
		tr\_tests\_ok&Number of tests passed  \\ 
		\hline 
		tr\_tests\_fail&Number of tests failed  \\ 
		\hline 
		tr\_tests\_run&Number of tests were run as part of this build  \\ 
		\hline 
		tr\_tests\_skipped&Number of tests were skipped or ignored in the build  \\ 
		\hline 
		tr\_testduration&Time it took to run the tests  \\ 
		\hline 
		gh\_test\_lines\_per\_kloc&Test density. Number of lines in test cases per 1000 gh\_sloc  \\ 
		\hline 
		gh\_test\_cases\_per\_kloc&Test density. Number of test cases per 1000 gh\_sloc  \\ 
		\hline 
		gh\_asserts\_cases\_per\_kloc&Assert density. Number of assertions per 1000 gh\_sloc  \\ 
		\hline 
		gh\_description\_complexity&If gh\_is\_pr, the total number of words in the pull request title and description  \\ 
		\hline 
		tr\_num\_jobs&How many jobs does this build have (lenght of tr\_jobs)  \\ 
		\hline 
		gh\_commits\_on\_files\_touched&Unique commits on the files included in the build from the moments the build was triggered and 3 months back  \\ 
		\hline 
		gh\_sloc&Number of executable production source lines of code, in the entire repository  \\ 
		\hline 
		tr\_setup\_time&Setup time for the Travis build to start  \\ 
		\hline 
		tr\_purebuildduration&Time it took to run the build (without Travis scheduling and provisioning the build)  \\ 
		\hline 
		git\_num\_committers&Number of people who committed to this project  \\ 
		\hline 
		tr\_ci\_latency&Latency included by Travis (scheduling, build pick-up, …)  \\ 
		\hline 
		gh\_is\_pr&Whether this build was triggered as part of a pull request on GitHub  \\ 
		\hline 
		tr\_tests\_ran&Whether tests ran in this build  \\ 
		\hline 
		tr\_tests\_failed&Whether tests failed in this build  \\ 
		\hline 
		gh\_by\_core\_team\_member&Whether this commit was authored by a core team member  \\ 
		\hline 
		tr\_duration&Overall duration of the build  \\ 
		\hline
	\end{tabular}
\end{table*}

\subsection{Evaluation Metrics}
Furthermore, we select an evaluation criteria to evaluate how well our model is performing in predicting build times from the learned dataset. This will also enable us to compare various algorithms to have an idea on which algorithm is performing better. The metric scores influence our options of what to pursue next to improve prediction accuracy. The metric used for the task is Root Mean Square Error (RMSE) and ${R}^2$ (R-Squared). 

RMSE reports the mean deviation of the predicted value from the original value. This gives us an idea of how well our algorithm is performing relative to the original value. The lower the RMSE, the better the prediction accuracy. The unit of RMSE is determined by the unit of the response variable, where in this case, it is in seconds, which is the unit of measurement for the total duration of a build job (\textit{tr\_duration}).

${R}^2$ gives us a measure of how much of the variation in predicted values is explained by the model. The values of ${R}^2$ ranges between 0 and 1. Values close to zero signify that a large proportion of variability in the result is unexplained by the model, while values close to one indicate that most of the variance in the result is accounted for by the model. We look for values closer to one to ensure the robustness of our model when encountered with new, unseen data.

\subsection{Initial Feature Scaling - Data Standardization}
The selected features for our prediction model are standardized to ensure that data values are in the same range. These features benefit some of the regression machine learning algorithms and instance based methods when evaluating the distance between points. Thus, we apply the center and scale standardization measure as a data pre-processing procedure.

\subsection{Rationale for Algorithm Selection}
We sample a set of linear and non-linear algorithms that work on regression problems to get a baseline performance on model accuracy.  In doing this, we use 10-fold cross validation with 3 repeats. This CV procedure splits the training set into 10-folds with each selected algorithm running 10 times on 90\% of the data, using the remaining 10\% to assess model performance. This process is repeated 3 times to produce an unbiased estimate of the algorithm performance. This also prevents over fitting the model by capturing noise from the data, which inherently leads to poor generalizability in predicting an out-of-sample build job time.

For this problem we sample the following Linear and Non-Linear Supervised Machine Learning algorithms to spot-check initial baseline results. For Linear models, we sample the following algorithms, Linear Regression (LR), Partial Least Squares (PLS), Penalized Linear Regression (GLMNET), and Least Angle Regression (LARS). For Non-Linear models, we sampled Classification and Regression Trees (CART), Support Vector Machines (SVM) with a radial basis function, k-Nearest Neighbors (KNN), and Neural Network (Nnet). We then spot-check a set of Ensemble methods such as Bagged Classification and Regression Trees (BCART), Random Forest (RF), Stochastic Gradient Boosting (SGB), and Cubist (CB) models.

The Linear models were selected because of their propensity to provide surprisingly good prediction results even when the inherent structural form of the data is non-linear. With enough data entries, linear models surprisingly perform well, sometimes out-performing or even equaling the performance of their non-linear counterparts on non-linear datasets. The non-linear models were selected because we understand the structure of the underlying relationships between features in the data to be non-linear. With a non-linear structure, a non-linear algorithm will be favored to perform better in prediction accuracy assessments. Finally, ensemble methods combine the outputs of various algorithms to get a better prediction score on unseen data. We sample a few of them here because ensemble methods are known to give good accuracy measures in various prediction tasks.

\subsection{Computational Tools}
Due to the computationally expensive nature of the project, we employed OpenStack, a cloud computing infrastructure as a service platform to run our algorithms to leverage the advantages of multicore parallelization. Our OpenStack configuration consisted of 20 CPUs which hosted a Linux distro of our computing tools.

We made use of the R statistical programming environment as the major tool for our analysis. R was selected as our tool of choice because of its robust, open-source machine learning packages. R has won large accolades in the area of predictive analytics. Many data scientists and machine learning engineers use R as their preferred tool of choice for predictive analytics \cite{vance}. We primarily made use of the caret package, among other key packages used for our work. Caret is short for ``Classification and Regression Training'' which contains a plethora of functions that simplify the process of training and testing machine learning models for regression and classification problems.

\section{Results}
\subsection{Benchmark}
To speed up the data processing time, we ran our algorithms on a subset of 10,000 records from the training set to get our initial baseline results on how the learning algorithms are performing. A seed was set to ensure reproducibility consistency. The results are shown in Table \ref{table-benchmark}.

\begin{table*}[t]
	\centering
	\caption{Benchmark Results}
	\label{table-benchmark}
	\begin{tabular}{llllllll}
		\hline
		\multicolumn{1}{|l|}{\textbf{RMSE}}     & \multicolumn{1}{l|}{\textbf{Min.}} & \multicolumn{1}{l|}{\textbf{1st Qu.}} & \multicolumn{1}{l|}{\textbf{Median}} & \multicolumn{1}{l|}{\textbf{Mean}} & \multicolumn{1}{l|}{\textbf{3rd Qu.}} & \multicolumn{1}{l|}{\textbf{Max.}} & \multicolumn{1}{l|}{\textbf{NA's}} \\ \hline
		\multicolumn{1}{|l|}{LM}                & \multicolumn{1}{l|}{4,682}         & \multicolumn{1}{l|}{4,953}            & \multicolumn{1}{l|}{5,384}           & \multicolumn{1}{l|}{5,839}         & \multicolumn{1}{l|}{5,535}            & \multicolumn{1}{l|}{11,410}        & \multicolumn{1}{l|}{0}             \\ \hline
		\multicolumn{1}{|l|}{PLS}               & \multicolumn{1}{l|}{4,695}         & \multicolumn{1}{l|}{4,956}            & \multicolumn{1}{l|}{5,142}           & \multicolumn{1}{l|}{5,769}         & \multicolumn{1}{l|}{5,484}            & \multicolumn{1}{l|}{11,310}        & \multicolumn{1}{l|}{0}             \\ \hline
		\multicolumn{1}{|l|}{GLMNET}            & \multicolumn{1}{l|}{4,666}         & \multicolumn{1}{l|}{4,961}            & \multicolumn{1}{l|}{5,120}           & \multicolumn{1}{l|}{5,755}         & \multicolumn{1}{l|}{5,478}            & \multicolumn{1}{l|}{11,320}        & \multicolumn{1}{l|}{0}             \\ \hline
		\multicolumn{1}{|l|}{LARS}              & \multicolumn{1}{l|}{4,682}         & \multicolumn{1}{l|}{4,954}            & \multicolumn{1}{l|}{5,368}           & \multicolumn{1}{l|}{5,825}         & \multicolumn{1}{l|}{5,518}            & \multicolumn{1}{l|}{11,400}        & \multicolumn{1}{l|}{0}             \\ \hline
		\multicolumn{1}{|l|}{CART}              & \multicolumn{1}{l|}{3,466}         & \multicolumn{1}{l|}{3,663}            & \multicolumn{1}{l|}{4,176}           & \multicolumn{1}{l|}{4,723}         & \multicolumn{1}{l|}{4,420}            & \multicolumn{1}{l|}{11,050}        & \multicolumn{1}{l|}{0}             \\ \hline
		\multicolumn{1}{|l|}{SVM}               & \multicolumn{1}{l|}{4,127}         & \multicolumn{1}{l|}{4,371}            & \multicolumn{1}{l|}{4,679}           & \multicolumn{1}{l|}{5,281}         & \multicolumn{1}{l|}{4,933}            & \multicolumn{1}{l|}{11,080}        & \multicolumn{1}{l|}{0}             \\ \hline
		\multicolumn{1}{|l|}{KNN}               & \multicolumn{1}{l|}{4,086}         & \multicolumn{1}{l|}{4,685}            & \multicolumn{1}{l|}{4,877}           & \multicolumn{1}{l|}{5,451}         & \multicolumn{1}{l|}{5,077}            & \multicolumn{1}{l|}{11,180}        & \multicolumn{1}{l|}{0}             \\ \hline
		\multicolumn{1}{|l|}{NNet}              & \multicolumn{1}{l|}{9,043}         & \multicolumn{1}{l|}{9,535}            & \multicolumn{1}{l|}{9,643}           & \multicolumn{1}{l|}{10,180}        & \multicolumn{1}{l|}{9,953}            & \multicolumn{1}{l|}{14,880}        & \multicolumn{1}{l|}{0}             \\ \hline
		\multicolumn{1}{|l|}{BCART}             & \multicolumn{1}{l|}{4,381}         & \multicolumn{1}{l|}{4,716}            & \multicolumn{1}{l|}{4,885}           & \multicolumn{1}{l|}{5,516}         & \multicolumn{1}{l|}{5,244}            & \multicolumn{1}{l|}{11,130}        & \multicolumn{1}{l|}{0}             \\ \hline
		\multicolumn{1}{|l|}{RF}                & \multicolumn{1}{l|}{2,309}         & \multicolumn{1}{l|}{3,043}            & \multicolumn{1}{l|}{3,580}           & \multicolumn{1}{l|}{4,145}         & \multicolumn{1}{l|}{4,190}            & \multicolumn{1}{l|}{10,650}        & \multicolumn{1}{l|}{0}             \\ \hline
		\multicolumn{1}{|l|}{SGB}               & \multicolumn{1}{l|}{2,753}         & \multicolumn{1}{l|}{3,336}            & \multicolumn{1}{l|}{3,682}           & \multicolumn{1}{l|}{4,374}         & \multicolumn{1}{l|}{3,956}            & \multicolumn{1}{l|}{10,780}        & \multicolumn{1}{l|}{0}             \\ \hline
		\multicolumn{1}{|l|}{CB}                & \multicolumn{1}{l|}{2,614}         & \multicolumn{1}{l|}{3,092}            & \multicolumn{1}{l|}{3,406}           & \multicolumn{1}{l|}{4,052}         & \multicolumn{1}{l|}{3,764}            & \multicolumn{1}{l|}{10,630}        & \multicolumn{1}{l|}{0}             \\ \hline
		\multicolumn{1}{|l|}{XGBOOST}           & \multicolumn{1}{l|}{3,094}         & \multicolumn{1}{l|}{3,647}            & \multicolumn{1}{l|}{3,959}           & \multicolumn{1}{l|}{4,693}         & \multicolumn{1}{l|}{4,709}            & \multicolumn{1}{l|}{10,770}        & \multicolumn{1}{l|}{0}             \\ \hline
		&                                    &                                       &                                      &                                    &                                       &                                    &                                    \\ \hline
		\multicolumn{1}{|l|}{\textbf{Rsquared}} & \multicolumn{1}{l|}{\textbf{Min.}} & \multicolumn{1}{l|}{\textbf{1st Qu.}} & \multicolumn{1}{l|}{\textbf{Median}} & \multicolumn{1}{l|}{\textbf{Mean}} & \multicolumn{1}{l|}{\textbf{3rd Qu.}} & \multicolumn{1}{l|}{\textbf{Max.}} & \multicolumn{1}{l|}{\textbf{NA's}} \\ \hline
		\multicolumn{1}{|l|}{LM}                & \multicolumn{1}{l|}{0.3053}        & \multicolumn{1}{l|}{0.5287}           & \multicolumn{1}{l|}{0.5602}          & \multicolumn{1}{l|}{0.5421}        & \multicolumn{1}{l|}{0.6003}           & \multicolumn{1}{l|}{0.6452}        & \multicolumn{1}{l|}{0}             \\ \hline
		\multicolumn{1}{|l|}{PLS}               & \multicolumn{1}{l|}{0.3149}        & \multicolumn{1}{l|}{0.5547}           & \multicolumn{1}{l|}{0.5731}          & \multicolumn{1}{l|}{0.5519}        & \multicolumn{1}{l|}{0.6009}           & \multicolumn{1}{l|}{0.6429}        & \multicolumn{1}{l|}{0}             \\ \hline
		\multicolumn{1}{|l|}{GLMNET}            & \multicolumn{1}{l|}{0.3178}        & \multicolumn{1}{l|}{0.5574}           & \multicolumn{1}{l|}{0.5732}          & \multicolumn{1}{l|}{0.5539}        & \multicolumn{1}{l|}{0.6017}           & \multicolumn{1}{l|}{0.6435}        & \multicolumn{1}{l|}{0}             \\ \hline
		\multicolumn{1}{|l|}{LARS}              & \multicolumn{1}{l|}{0.3058}        & \multicolumn{1}{l|}{0.5349}           & \multicolumn{1}{l|}{0.5643}          & \multicolumn{1}{l|}{0.5441}        & \multicolumn{1}{l|}{0.6003}           & \multicolumn{1}{l|}{0.6451}        & \multicolumn{1}{l|}{0}             \\ \hline
		\multicolumn{1}{|l|}{CART}              & \multicolumn{1}{l|}{0.3445}        & \multicolumn{1}{l|}{0.7213}           & \multicolumn{1}{l|}{0.7351}          & \multicolumn{1}{l|}{0.7127}        & \multicolumn{1}{l|}{0.7878}           & \multicolumn{1}{l|}{0.8039}        & \multicolumn{1}{l|}{0}             \\ \hline
		\multicolumn{1}{|l|}{SVM}               & \multicolumn{1}{l|}{0.3331}        & \multicolumn{1}{l|}{0.6382}           & \multicolumn{1}{l|}{0.6657}          & \multicolumn{1}{l|}{0.6317}        & \multicolumn{1}{l|}{0.6888}           & \multicolumn{1}{l|}{0.7292}        & \multicolumn{1}{l|}{0}             \\ \hline
		\multicolumn{1}{|l|}{KNN}               & \multicolumn{1}{l|}{0.3326}        & \multicolumn{1}{l|}{0.6068}           & \multicolumn{1}{l|}{0.6339}          & \multicolumn{1}{l|}{0.6066}        & \multicolumn{1}{l|}{0.66}             & \multicolumn{1}{l|}{0.7315}        & \multicolumn{1}{l|}{0}             \\ \hline
		\multicolumn{1}{|l|}{NNet}              & \multicolumn{1}{l|}{NA}            & \multicolumn{1}{l|}{NA}               & \multicolumn{1}{l|}{NA}              & \multicolumn{1}{l|}{NaN}           & \multicolumn{1}{l|}{NA}               & \multicolumn{1}{l|}{NA}            & \multicolumn{1}{l|}{30}            \\ \hline
		\multicolumn{1}{|l|}{BCART}             & \multicolumn{1}{l|}{0.3384}        & \multicolumn{1}{l|}{0.5784}           & \multicolumn{1}{l|}{0.6309}          & \multicolumn{1}{l|}{0.5942}        & \multicolumn{1}{l|}{0.6448}           & \multicolumn{1}{l|}{0.6864}        & \multicolumn{1}{l|}{0}             \\ \hline
		\multicolumn{1}{|l|}{RF}                & \multicolumn{1}{l|}{0.3989}        & \multicolumn{1}{l|}{0.7474}           & \multicolumn{1}{l|}{0.8091}          & \multicolumn{1}{l|}{0.7742}        & \multicolumn{1}{l|}{0.8436}           & \multicolumn{1}{l|}{0.9143}        & \multicolumn{1}{l|}{0}             \\ \hline
		\multicolumn{1}{|l|}{SGB}               & \multicolumn{1}{l|}{0.3787}        & \multicolumn{1}{l|}{0.7597}           & \multicolumn{1}{l|}{0.7834}          & \multicolumn{1}{l|}{0.7489}        & \multicolumn{1}{l|}{0.8102}           & \multicolumn{1}{l|}{0.8779}        & \multicolumn{1}{l|}{0}             \\ \hline
		\multicolumn{1}{|l|}{CB}                & \multicolumn{1}{l|}{0.4}           & \multicolumn{1}{l|}{0.7905}           & \multicolumn{1}{l|}{0.8145}          & \multicolumn{1}{l|}{0.7808}        & \multicolumn{1}{l|}{0.8433}           & \multicolumn{1}{l|}{0.8892}        & \multicolumn{1}{l|}{0}             \\ \hline
		\multicolumn{1}{|l|}{XGBOOST}           & \multicolumn{1}{l|}{0.3795}        & \multicolumn{1}{l|}{0.69}             & \multicolumn{1}{l|}{0.7629}          & \multicolumn{1}{l|}{0.7134}        & \multicolumn{1}{l|}{0.7828}           & \multicolumn{1}{l|}{0.8455}        & \multicolumn{1}{l|}{0}             \\ \hline
	\end{tabular}
\end{table*}

From the results provided in Figure 1, Cubist (CB) model has the lowest RMSE of 4,052 seconds, followed by Random Forest (RF) with 4,145 seconds. While Cubist (CB) model and Stochastic Gradient Boosting (SGB) also have the highest and second-highest ${R}^2$ value of 0.7808 and 0.7742 respectively. An average baseline difference of 4,052 seconds is the accuracy measure to beat in order to improve the model. From experience this CV RMSE value is likely to increase when tested on unseen data. We explored other predictive analytic techniques to see if we can get a better prediction model with a lower RMSE and a higher ${R}^2$.

\subsection{Drop Highly Correlated Features}
In machine learning practice, it is observed that features with high correlation can have an adverse effect on the predictive model accuracy \cite{domingos2012few}. Hence, we employ this technique to prune out features that are highly correlated. To do this, we set a correlation cut-off score of 0.70 (i.e. values about 0.70 or below -0.70) to indicate highly correlated variables. We employ the findCorrelation() function of the Caret package to find and remove highly correlated features. We use the subset of 10,000 records from the training set to compute the correlation matrix. The features \textit{gh\_src\_files}, \textit{tr\_tests\_ok}, \textit{gh\_test\_cases\_per\_kloc}, and \textit{gh\_test\_lines\_per\_kloc} had a correlation index greater than 0.70 or less than -0.70, hence they are removed from the model. This reduced the number of features to 31.

However, we did not get a noticeable improvement in the prediction accuracy by taking out this attributes. In Table \ref{table-rmse} are the mean values of the RMSE of ${R}^2$ error measures.

\begin{table}[h]
	\centering
	\caption{RMSE of ${\textbf{R}}^\textbf{2}$ Error Measures}
	\label{table-rmse}
	\begin{tabular}{|l|l|l|}
		\hline
		\textbf{Algorithm} & \textbf{RMSE} & \textbf{R2} \\ \hline
		LM                 & 5,767         & 0.5522      \\ \hline
		PLS                & 5,768         & 0.5522      \\ \hline
		GLMNET             & 5,121         & 0.5538      \\ \hline
		LARS               & 5,767         & 0.5522      \\ \hline
		CART               & 4,754         & 0.7076      \\ \hline
		SVM                & 5,313         & 0.6271      \\ \hline
		KNN                & 5,466         & 0.6048      \\ \hline
		NNet               & 10,180        & NA          \\ \hline
		BCART              & 5,474         & 0.5989      \\ \hline
		RF                 & 4,224         & 0.7659      \\ \hline
		SGB                & 4,416         & 0.7446      \\ \hline
		CB                 & 4,090         & 0.7767      \\ \hline
		XGBOOST            & 3,928         & 0.7134      \\ \hline
	\end{tabular}
\end{table}

\subsection{Recursive Feature Selection (RFE)}
\begin{figure}[htpb]
	\centering
	\caption{Recursive Feature Selection}
	\label{figure-rfs}
	\includegraphics[height=2in]{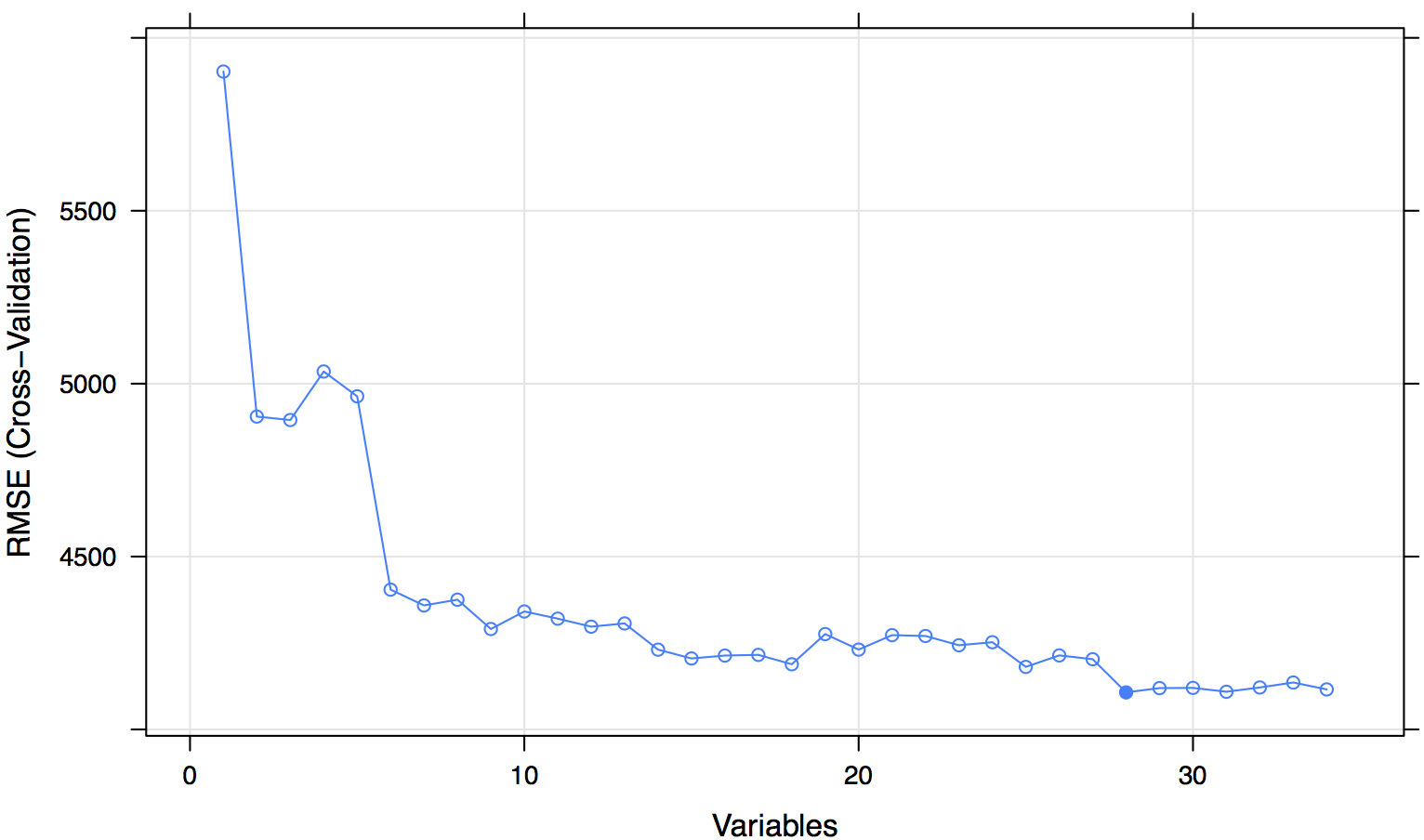}
\end{figure}

RFE is an automatic method of selecting features for a predictive model based on their relative importance. RFE is defined as a wrapper method, this is because of the way it samples the features, as it analyzes the interactions between increasing subsets of features to determine the relative impact of each feature in the presence of others. This can also be viewed as a brute-force method; it is computationally expensive. We used the random forest implementation of RFE to search and identify the best feature space that will improve our model accuracy. The results of implementing this automatic selection method is shown in Figure \ref{figure-rfs}.

From Figure \ref{figure-rfs}, we hope to achieve a slightly better model by reducing the feature space to 28 variables. However, the mean prediction difference was not statistically significant.

\subsection{Boruta Feature Selection Method}
We also employed another wrapper technique called Boruta, to see if we can squeeze out an improvement on our prediction accuracy on the CV dataset. Boruta is another automatic wrapper method that uses random forest in its algorithm to determine the relative importance of features with respect to the response variable \cite{kursa2010feature}. We achieved similar results to the benchmark after running this algorithm and testing the new dataset of 29 features across our selected learning algorithms.

\subsection{Applying Box-Cow Power Transforms}
We applying the Box-cow power transforms, this further normalizes our data features to approximate a Gaussian distribution. This data transformation technique is heuristically known to improve prediction accuracy across various ``linear'' machine learning models that perform better on normalized data \cite{proietti2013does}. However, this transform has its biggest improvements across our linear models. It has no effect on on overall accuracy threshold because linear models are generally performing very poorly for this problem as previous results have shown.

\subsection{Using Principal Component Analysis}
Principal Component Analysis is another data pre-processing technique that we implemented to extract important features from our dataset. This method is relevant when you have a high dimensional dataset, and you need to scale down the feature set to the ones that contain just the information you need to optimize prediction accuracy\cite{pca-analytics-vidhya}. Before applying principal component analysis, we first normalized our features (i.e. applying center and scale operations) to have them on the same scale. Unfortunately, the results of applying PCA did not yield any significant improvement on our CV accuracy. The results are shown in Figure \ref{figure-pca} below.

\begin{figure}[htpb]
	\centering
	\caption{Principal Component Analysis}
	\label{figure-pca}
	\includegraphics[height=3.5in]{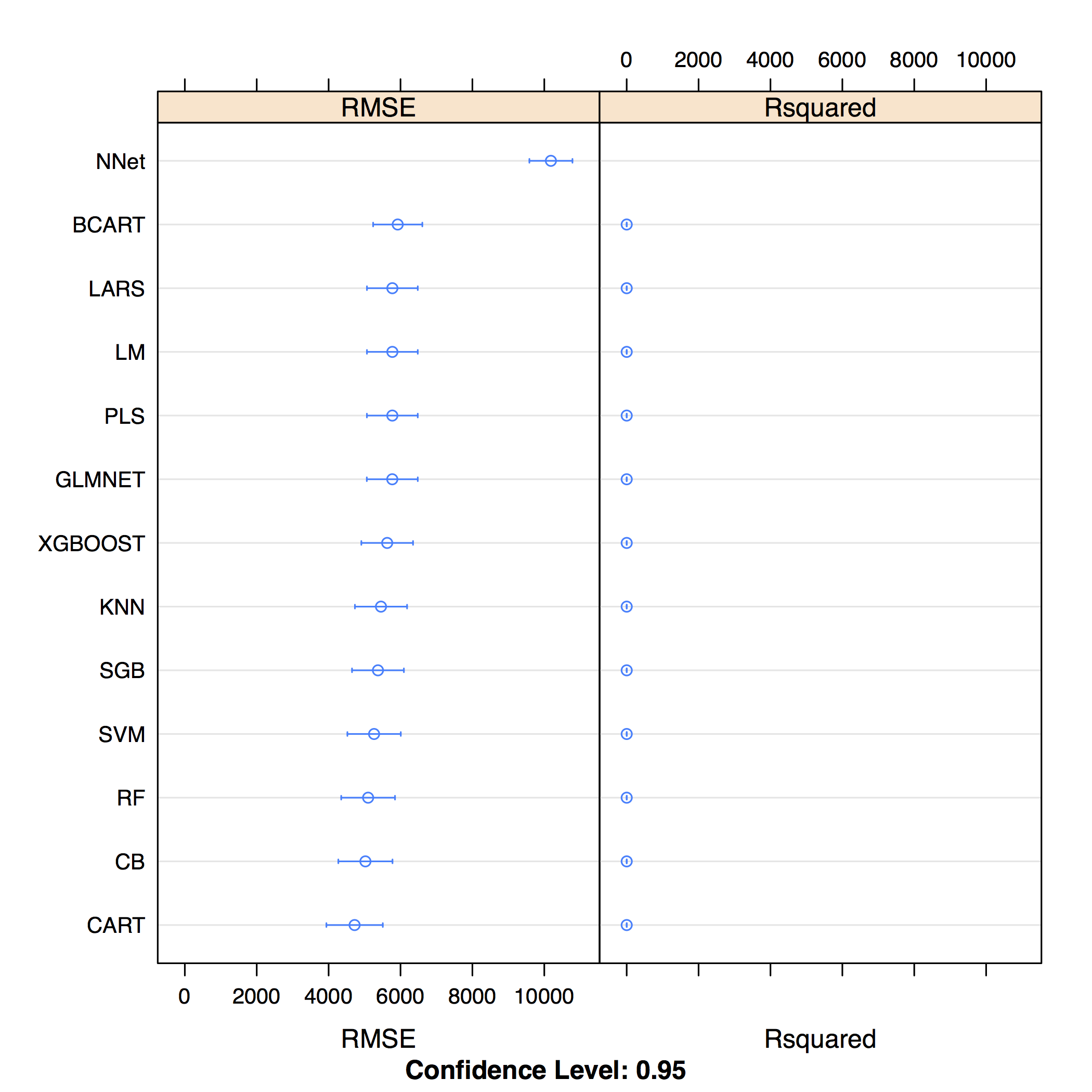}
\end{figure}

\subsection{Test Set Accuracy on Selected Models}
Finally, we applied our baseline models on unseen data (i.e. the test set) to see how our model performs on out-of-sample data points. Usually, we expect our test set errors to predict slightly worse than our CV set error estimates. This is because our CV set can sometimes give us overly-optimistic prediction values.

The test set was constructed by sampling 10,000 unseen records from the original 791,310 records. We used each of the final models to predict the build duration of the test data. The model estimates were compared to the original \textit{tr\_duration} value of the test set using RMSE and ${R}^2$ as evaluation metrics.

\begin{figure}[htpb]
	\centering
	\caption{Test Set Prediction - ${R}^2$}
	\label{figure-final-r2}
	\includegraphics[height=1.8in]{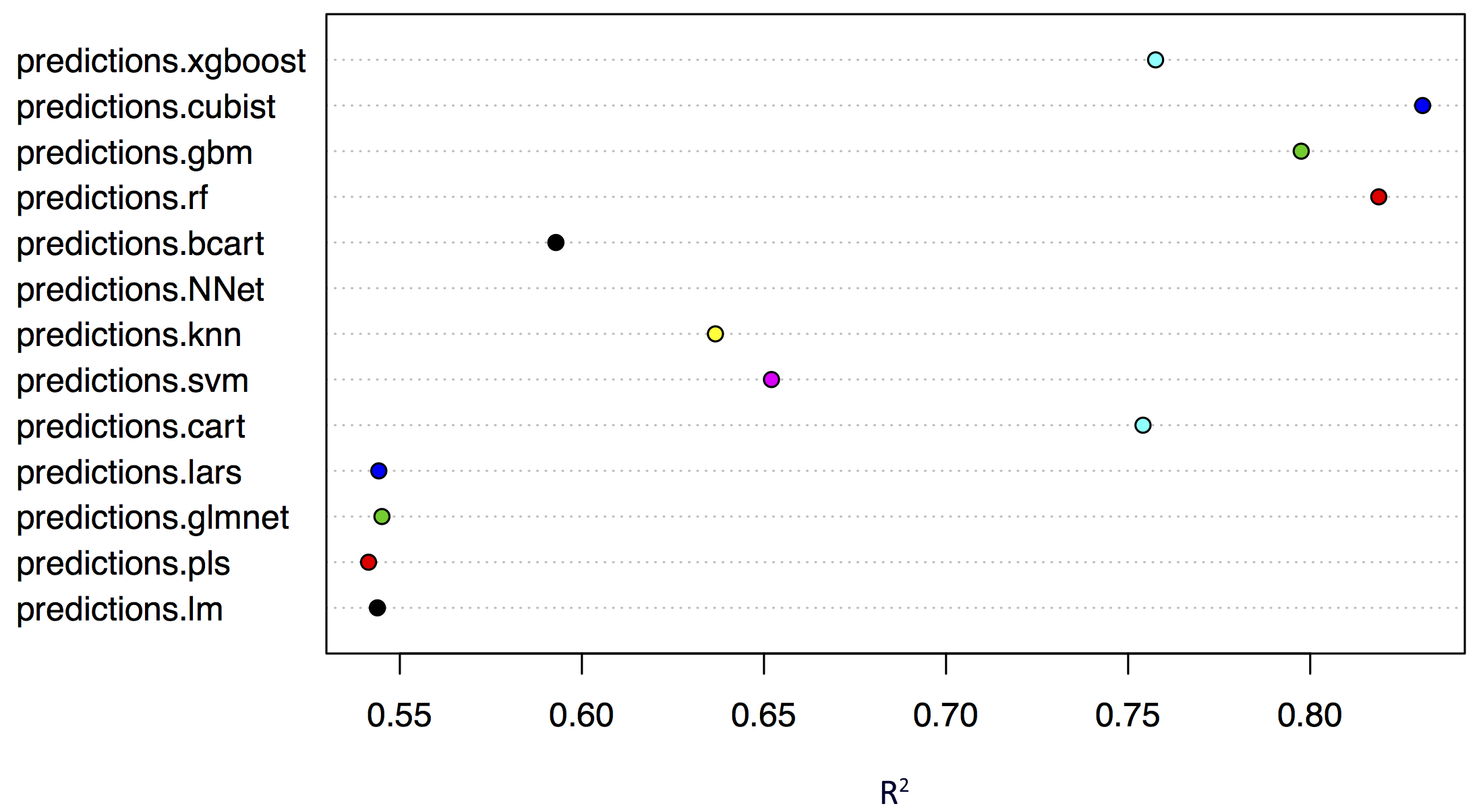}
\end{figure}

\begin{figure}[htpb]
	\centering
	\caption{Test Set Prediction - RMSE}
	\label{figure-final-rmse}
	\includegraphics[height=1.8in]{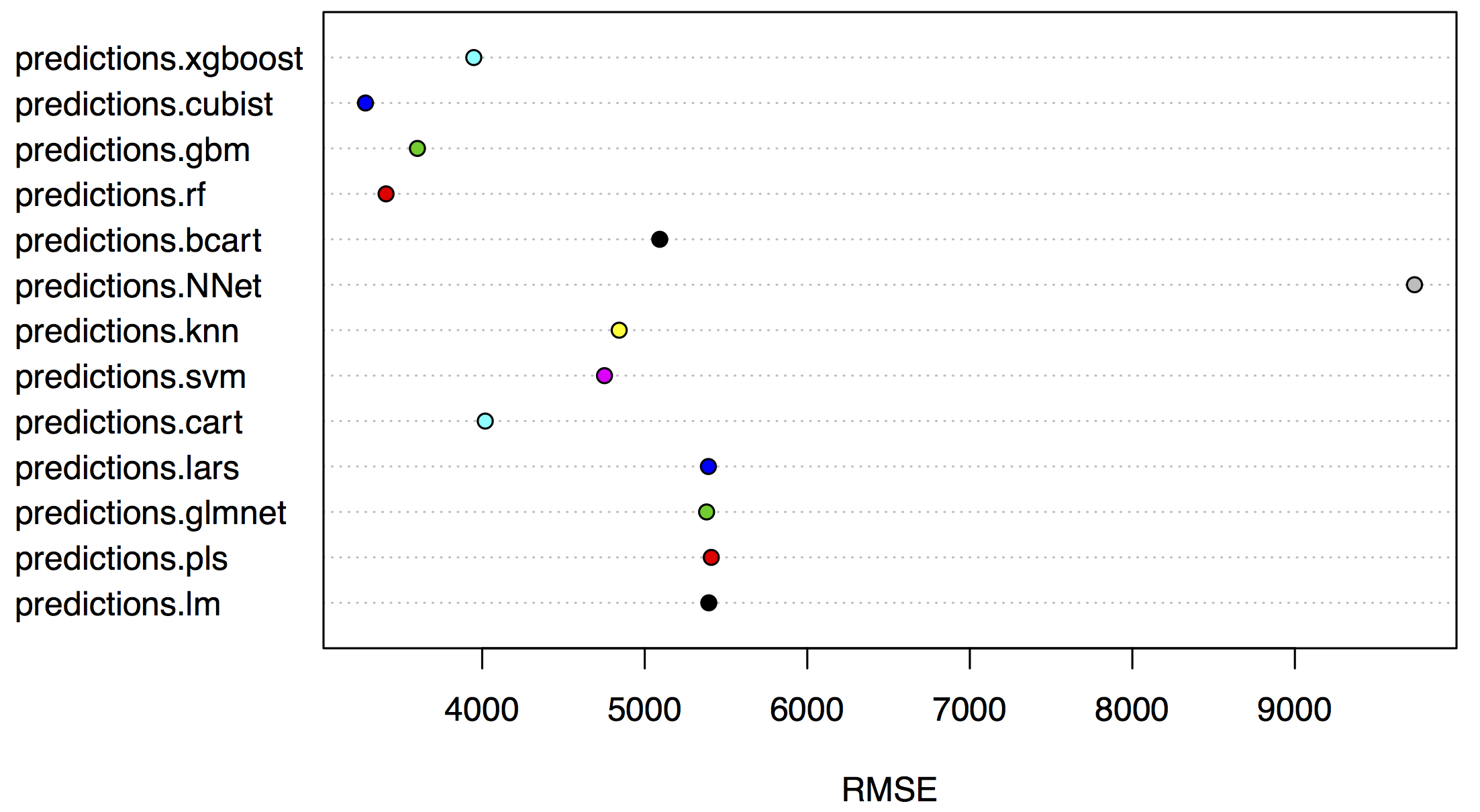}
\end{figure}

Table \ref{final-test-set} provides a tabular summary of the test-set prediction accuracy. This is the result of predicting the different models on unseen data. As expected from the results on the CV prediction accuracies, Cubist (CB) and Random Forest (RF) are outperforming the others with a lower RMSE and higher ${R}^2$. This ${R}^2$ metric is particularly very encouraging, because it shows us that a high percentage of variance in the prediction accuracies is accounted for by the model. Figure \ref{figure-final-rmse} \& \ref{figure-final-r2} presents a graphical view of the test-set RMSE and ${R}^2$ in a dotplot.

\begin{table}[h]
	\centering
	\caption{Metric Performance on Test Set for our Final Models}
	\label{final-test-set}
	\begin{tabular}{|l|l|l|}
		\hline
		\textbf{Algorithm} & \textbf{RMSE} & \textbf{${R}^2$} \\ \hline
		LM                 & 5,394         & 0.544      \\ \hline
		PLS                & 5,409         & 0.541      \\ \hline
		GLMNET             & 5,380         & 0.545      \\ \hline
		LARS               & 5,391         & 0.544      \\ \hline
		CART               & 4,018         & 0.754      \\ \hline
		SVM                & 4,752         & 0.652      \\ \hline
		KNN                & 4,842         & 0.637      \\ \hline
		NNet               & 9,736        & NA          \\ \hline
		BCART              & 5,092         & 0.593      \\ \hline
		RF                 & 3,408         & 0.819      \\ \hline
		SGB                & 3,601         & 0.798      \\ \hline
		CB                 & 3,281         & 0.831      \\ \hline
		XGBOOST            & 3,948         & 0.756      \\ \hline
	\end{tabular}
\end{table}

\section{Discussion}
\subsection{Implications of Study}
Wallace et. al \cite{wallace2004software} suggests that as the number of developers increases, the project size and complexity also widens. This is turn escalates the wait time of a build job. Also, frequent integration follows that a minimal number of changes in software artifacts (e.g. lines of code changed, files added, files deleted, files modified, etc.) are constantly integrated into the main code base.

Developers tend to lose focus and productivity while waiting for code to build \cite{kai:context}. This has a detrimental effect to the idea of continuous integration which advocates for frequent builds to commit changes to the central code base fast and early. However, as the projects expands, frequent integration can become a huge impedance to productivity. For example, if a build takes approximately 1 hour or more, integrating more than once a day can become a setback to developer speed and efficiency.

From the related work, numerous techniques have been discussed to balance the trade-off between build waiting time and the need to continuously integrate code. Our research project comes in the middle to further balance this trade-off. We developed a predictive model to approximate the build time of a build job in a CI environment.

\textbf{Software Developers.} 
When the approximate time for a build is known, a developer can be more intentional with how they spend their time. In turn, the developer will improve their efficiency and productivity. For example, if a build is expected to take a long time, then the developer may move onto other tasks such as responding to e-mails or code reviews. As a whole, the overall quality of their project will improve.

\textbf{Project Managers.}
Knowing the approximate build time beforehand can be advantageous to project managers (and management in general). Project managers are able to strategize on how best to manage the CI process to balance the anticipated build wait time versus the need to continuously commit changes to the central code base. This pre-knowledge will be particularly crucial for large projects. Project managers can explore different variables to ascertain the best setup for a build job that will keep the build wait time within an acceptable level, given the peculiar circumstance of the project. Teams can be aware of the minimum amount of changes that must be implemented before a build job is triggered.

\textbf{Software Organizations.}
Software organizations looking to adopt CI can make use of the prediction model. As organizations transition to CI practices, they may begin to approximate their build times. Making process changes that favourably reduce build times may be easier when CI practices are not fully implemented within an organization. The lack of constraints from a CI environment may allow organizations to quickly make changes to reduce build times before CI is eventually implemented.

\textbf{Researchers.}
Finally, other researchers can use and modify the prediction model to further our understanding of build times. Researchers can further study the variables and relationships affecting build times. The prediction model could be expanded to include a dataset other than TravisTorrent.

Our research has a direct and immediate relevance to the industry, and further strengthens the concept of continuous integration, which consequently gives rise to continuous delivery in the widely industry embraced agile development model.

\begin{figure*}[htpb]
	\centering
	\caption{Build Time Predictor App}
	\label{app-screenshot}
	\includegraphics[width=\textwidth]{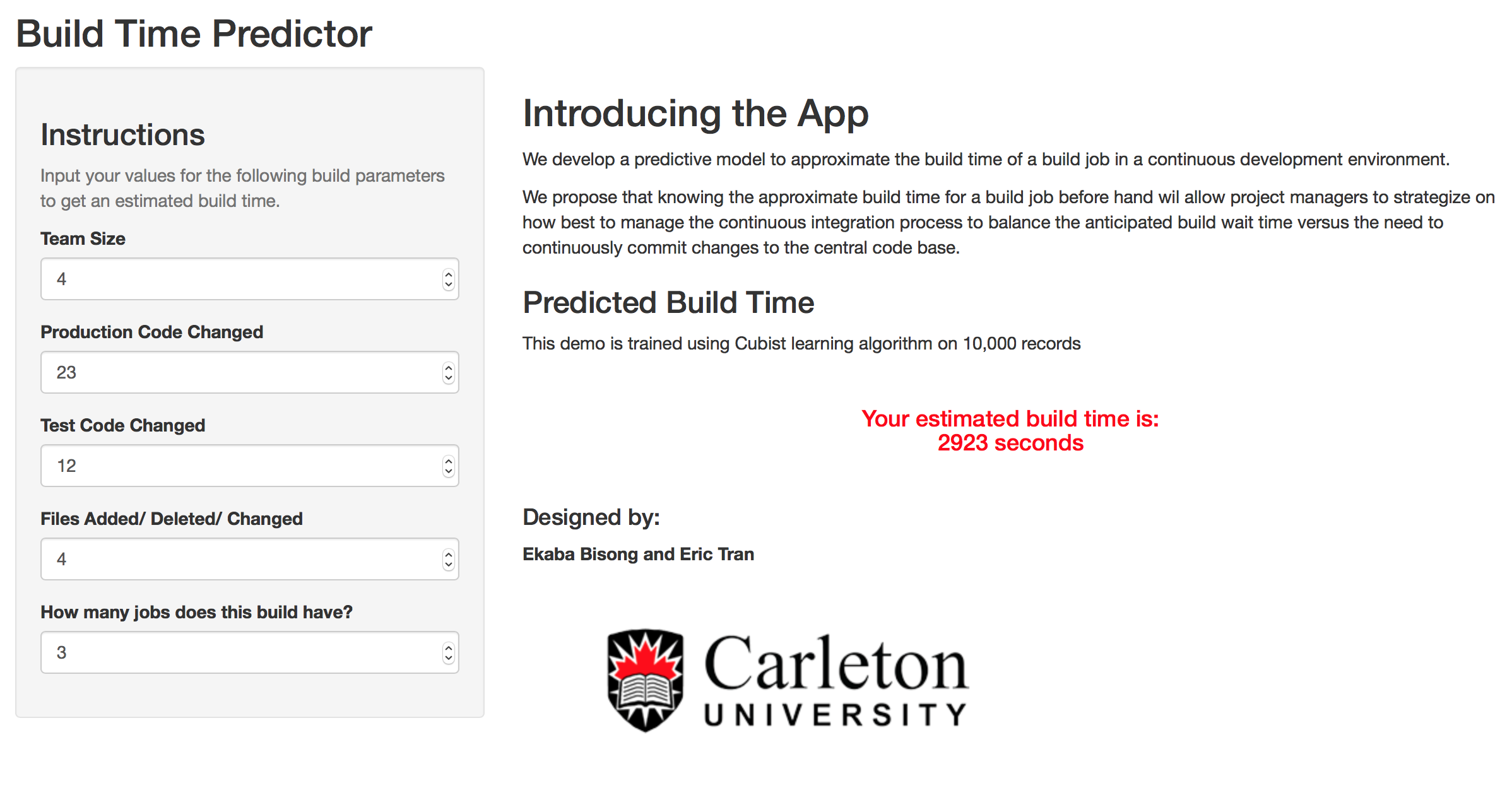}
\end{figure*}

\subsection{Limitations of Study}
Our study had several limitations that mitigated further enquiry into the improvement of our predictive model. The main obstacle was the sheer computation power required to perform our analysis. The TravisTorrent dataset contains approximately 1.76GB of data with more than 2 million observations. 

Additionally, in this problem, we considered a variety of learning algorithms, which comprised of a combination of linear, non-linear, and ensemble methods. Some of the learning algorithms are computationally expensive: this is especially true of kernel methods such as support vector machines and ensemble methods like random forest and stochastic gradient boosting. These algorithms took over 8 hours each for a single run. This is taking into consideration the fact that our original dataset was sub-sampled from 1,846,396 records to 10,000 records for the training set.

We ran into issues with R, the computational tool that we employed for our analysis. We encountered a lot of technical issues when running R on our dedicated OpenStack cluster of 20 cores; thus our tasks were prone to frequent crashing. We were forced to restart our R sessions multiple times, sometimes even after running an algorithm for over 8-10 hours. R in our experience for this research project does not have a very mature parallelization framework. We believe that at this point in time, R is unsuitable for multi-core, high performance data analytic computing.

\subsection{Reproducibility}
The data and code used in this research has been made publicly available. This is to enhance reproducibility and further mining enquiry to improve on our presented results. We have made this data available in the spirit of open research and collaborative mining enquiry. The authors are not perfect, we however hope that constructive criticisms, revisions and remodeling can be made on our work.

\subsection{Further Recommendations}
Base R does not scale well as a tool for parallel computing with big data. Hadoop would do a better job in this area, as indeed this is what Hadoop was primarily built for. Furthermore, R operates on data from RAM, and it can run out of RAM space quickly when carrying out expensive computations. On the other hand, Hadoop works with data stored on disks which is usually more in supply. Due to time constraints we could not fully explore working with Hadoop and its MapReduce operations to crunch our data.

Typically, the predictive accuracy of a model is improved when more data is available. We could not take advantage of the copious amounts of data available to train our model for obvious reasons. Hadoop is strongly recommended by the authors for further exploration.

\subsection{Demo App to Motivate Integration as an Industry Tool}
In conclusion, we hope the features space can be further simplified to a set of variables that can be implemented as a tool in industry to predict the build time of various GitHub projects running Travis as a CI platform. To motive that desire, we have created a sample application to further communicate that thought. The application can be viewed at \url{https://dvdbisong.shinyapps.io/BuildTimePredictor/}. This tool can be implemented as an IDE plugin or as an add-on in a planning/scheduling software for developers. Figure \ref{app-screenshot} shows a screenshot of the application page.

In building the sample application, we used a trained Cubist model. We formulated a sample test by receiving as input the values of team size, lines of production code changed, test code changed, files added, deleted or changed and the number of jobs contained in the build. The remaining 30 variables were estimated by using the means of the values in the test data sample. Hopefully a future study can reduce the feature space to a small set of relevant features that will minimize RMSE and maximize the ${R}^2$ error metric.

\section{Acknowledgments}
We would like to thank Professor Olga Baysal for teaching COMP 5900: Mining Software Repositories, for which we were fortunate to have taken. Through weekly presentations as well as back-and-forth discussions, we gained a lot of insight about the software development industry. We believe that there is much more that can be discovered!

We would also like to thank Andrew Pullin for his assistance in helping us set up our computing environment. Andrew laid out some best practices and helped us set up our environment OpenStack.

\end{document}